\documentclass[11pt,aaspp4]{aastex}

\newcommand{\HI}{{\ion{H}{1}}}
\newcommand{\matHI}{\rm H{\hskip 0.02cm\scriptscriptstyle I}}

\newcommand{\kms}{$\,$km$\,$s$^{-1}$}
\newcommand{\atms}{atoms cm$^{-2}$}

\newcommand{\etal}{{\sl et al.}}
\newcommand{\kmsMp}{km s$^{-1}$ Mpc$^{-1}$}
\newcommand{\mJybeam}{mJy beam$^{-1}$}
\newcommand{\msun}{{$M_\odot$}}
\newcommand{\lsun}{{$L_\odot$}}
\input{psfig}

\begin{document}

\title{Extended \HI\ disks in dust--lane elliptical 
galaxies\footnote{Based on observations with the Australia Telescope Compact
Array (ATCA) which is funded by the Commonwealth of Australia for operation as
a National Facility managed by CSIRO}}

\author{Thomas A. Oosterloo}  
\affil{ASTRON, PO Box 2, 7990 AA,
Dwingeloo, The Netherlands}
\email{oosterloo@astron.nl}
\author{Raffaella Morganti}
\affil{ASTRON, PO Box 2, 7990 AA,
Dwingeloo, The Netherlands}
\email{morganti@astron.nl}
\author{Elaine M. Sadler}
\affil{School of Physics, University of Sydney, NSW\,2006, Australia}
\email{ems@physics.usyd.edu.au}
\author{Daniela Vergani}
\affil{Dipartimento di Astronomia, Universit\`a di Bologna, via Zamboni 33,
I-40126 Bologna, Italy}
\author{Nelson Caldwell}
\affil{Smithsonian Institution, Steward Observatory, 949 N Cherry Avenue,
Tucson Arizona 85719, USA}  
\email{caldwell@flwo99.sao.arizona.edu}

%\vskip0.5cm
%\centerline{$ $Revision: 1.25 $ $}
%\centerline{$ $Date: 2001/07/11 01:29:58 $ $}
%\centerline{$ $Author: toosterl $ $}

\begin{abstract}
  
  We present the results of \HI\ observations of five dust-lane ellipticals
  with the Australia Telescope Compact Array.  Two galaxies (NGC~3108 and
  NGC~1947) are detected, and sensitive upper limits are obtained for the
  other three. In the two detected galaxies, the \HI\ is distributed in a
  regular, extended and warped disk-like structure of low surface brightness.
  Adding data from the literature, we find that several more dust-lane
  ellipticals have regular \HI\ structures.  This \HI\ is likely to be a
  remnant of accretions/mergers which took place a considerable time ago, and
  in which a significant fraction of the gas survived to form a disk.  The
  presence of regular \HI\ structures suggests that some mergers lead to
  galaxies with extended low surface brightness density gas disks. These gas
  disk will evolve very slowly and these elliptical galaxies will remain gas
  rich for a long period of time.

One of the galaxies we observed (NGC~3108) has a very large amount of neutral
hydrogen ($M_{\matHI} = 4.5 \times 10^9$ \msun; $M_{\matHI}/L_B \sim 0.09 $),
which is very regularly distributed in an annulus extending to a radius of
$\sim$6 $R_{\rm eff}$.  The kinematics of the \HI\ distribution suggest that
the rotation curve of NGC 3108 is flat out to at least the last observed
point.  We estimate a mass-to-light ration of $M/L_B \sim 18$
$M_\odot/L_{B,\odot}$ at a radius of $\sim$6 $R_{\rm eff}$ from the centre.

Several of the galaxies we observed  have an unusually low gas--to--dust ratio
$M_{\matHI}/M_{\rm  dust}$, suggesting  that   their cold ISM, if   present as
expected  from the presence  of dust, may be  mainly  in molecular rather than
atomic form.
\end{abstract}

\section{Introduction}

Theoretical  models of  galaxy formation and  evolution have  pointed out that
differences between gas content and gas supply are a  key factor in explaining
differences between  different galaxies  (e.g.\ Kauffman  1996).  At the  same
time, \HI, FIR  and CO observations of  elliptical galaxies  have demonstrated
that many of  these objects have  an active and  interesting cold interstellar
medium,  often  qualitatively similar to   that  observed in spirals, although
usually  (but  not always)  in  much smaller quantity  (see   Knapp 1998 for a
review).  The limits in sensitivity of  the available instruments is still the
main limitation for studying the cold ISM in elliptical galaxies in a detailed
an unbiased way.

It is widely believed that when neutral hydrogen  is observed in an elliptical
galaxy it is the result of a recent accretion  of a gas-rich companion galaxy.
Indeed, gas-rich elliptical galaxies are often classified as `peculiar'.  This
implies that by studying such galaxies one is considering only a subset of the
whole population of early-type galaxies,  namely those for  which it is likely
that   some  interaction/accretion has  occurred.  But   when investigating the
evolution of the whole group of early-type galaxies, it could be important not
to restrict samples to `pure' ellipticals with no optical peculiarities, since
gas-rich systems  may represent an important   phase in the evolution  of many
early-type galaxies and  give us  important information  on the formation  and
evolution of  these systems.   At high redshift,  similar processes  will have
been more common.

The correlation between optical `peculiarities' and increased probability of
finding \HI\ can be interpreted (van Gorkom \& Schiminovich 1997) either as an
effect of the {\sl evolutionary stage} of these galaxies (accretion of \HI\ is
a normal phase in the evolution of these galaxies) or as an effect of {\sl
  environment} (the fraction of peculiar ellipticals in the Revised
Shapley-Ames Catalogue goes from 5\% in clusters to almost 50\% in the field)
or (perhaps more likely) as a combination of the two.  If the accretion of gas
is a normal event in the evolution of many or most early-type galaxies, this
raises questions like how much of the observed ISM is due to small accretion
events and to what extent parallels with spiral galaxies exist, whether in
major mergers a significant fraction of the gas remains and, if so, how long
does the ISM survive and will it settle in a disk-like structure.

Recently, there have been several studies of major mergers (i.e.\ mergers of
two disk galaxies) and the fate of gas in these systems (Hibbard \& van Gorkom
1996; Hibbard \& Mihos 1995). These studies have shown the importance of the
late infall of gas, possibly leading to extended, long--lived gas--rich
structures in elliptical galaxies.  Indeed, an increasing number of early-type
galaxies are known that have a surprisingly large amount of \HI\ distributed
in quite {\sl regular} structures (i.e.\ disks or rings, e.g., in IC 2006, NGC
4278, NGC 2974, NGC~5266, IC 5063).  The regular structure of these disks and
rings means that they are relatively old, indicating that some early--type
galaxies have a long-lived ISM. In such galaxies, the cold ISM is not a
`peculiarity', but a fundamental constituent.

Apart from  the  origin and evolution  of  the neutral  hydrogen in early-type
systems, studying   \HI\ in these objects   and in  particular  in dust--lane
ellipticals,  may give accurate estimates   of the mass  distribution at large
radii in these  galaxies.  Bertola  et al.\  (1993) combined  $M/L$
values derived from \HI\  data with estimates for  the central $M/L$ (based on
optical measurements from ionized     gas disks) to   show evidence    for  an
increasing   $M/L$ with radius,  supporting  the  idea  of a  dark halo around
ellipticals.  However estimates of $M/L$ from \HI\ data are only available for
a handful of well--studied ellipticals   (e.g.\ IC~2006, Franx et al.\   1994;
NGC~1052, van Gorkom  et al.\  1986; NGC~4278, Raimond  et al.\  1981 and Lees
1994; NGC~5266, Morganti   et al.\ 1997).   It is  important to  increase the
number of galaxies  for which a similar analysis  using \HI\ data can be done.
This requires   not only finding  ellipticals with  \HI,  but finding those in
which the \HI\  has settled into  a regular disk.  For dust-lane  ellipticals,
regular large-scale  dust could indicate  the presence  of a regular  disk and
this is particularly valuable if  we want to determine  the intrinsic shape of
the galaxy and the mass--to--light ratio ($M/L$).

In this paper we present new observations of neutral hydrogen in dust--lane
elliptical galaxies.  The presence of dust lanes is often considered a
`peculiarity' in an elliptical galaxy.  Thus, for the reasons mentioned above,
dust--lane ellipticals are good targets in which to search for neutral
hydrogen.  Systematic optical searches for dust in elliptical galaxies
(Goudfrooij \& de Jong 1995; van Dokkum \& Franx 1995) have found dust lanes
and patches in the inner regions of about 50\% of the observed galaxies.
Recent HST studies have shown that a large fraction (40%) of ellipticals have
dust of some sort visible in optical images (Rest et al.\ 2001); either in
filamentary form, nuclear disks or more large scale disks.  In this paper we
study five galaxies with large-scale dust lanes (Centaurus~A-like), mainly
taken from the compilation of Sadler \& Gerhard (1985): NGC~1947, NGC~3108,
ESO~263--G48, NGC~7049, NGC~7070A.  Table~1 summarizes the main properties of
these five galaxies.  \HI\ observations of these galaxies were made with the
Australia Telescope Compact Array (ATCA) as part of a long term project to
increase the number of early--type galaxies with detailed \HI\ images
(Morganti et al.\ 1997, 1998, Oosterloo et al.\ 1999a,b, 2001; Sadler et al.\
2000).  We present the results from these new observations in \S3, and in \S4
we discuss them in the context of existing \HI\ data on dust-lane ellipticals
in the literature.

In this paper we use $H_{\circ} = 50$ \kmsMp.

\section{Observations and Data Reduction}

All five galaxies  were  first observed with   the most compact  configuration
available with the ATCA  (375\,m array).  The  two detected galaxies (NGC~1947
and NGC~3108, see below) were then observed  with other ATCA configurations to
image the emission  in more detail.  Table~2 summarizes the \HI\ observations.
We  used  a 16 MHz band   with 512 velocity  channels   centred on the optical
velocity   of  each galaxy.   The   final  velocity resolution (after  Hanning
smoothing the data) is $\sim$12 km s$^{-1}$.

We observed each galaxy for about 12 hours in each  run (with the exception of
NGC~7070A), with  calibrators  observed every  hour to  monitor gain and phase
changes.  The flux density scale was set by observations of PKS 1934--638, for
which we adopted a flux density of 14.9 Jy at 1400  MHz.  This source was also
used as bandpass calibrator.

The  spectral data  were calibrated with   the MIRIAD package  (Sault  et al.\
1995), which has  several  features particularly suited for   ATCA data.   The
continuum subtraction  was also done in MIRIAD,  by using a linear fit through
the line--free channels of   each visibility record  and subtracting  this fit
from all  the  frequency  channels  (i.e.\   UVLIN).   An interference   spike
generated by the ATCA data acquisition system is present at 1408~MHz, but this
does not compromise the interpretation of the data.

For the objects which were observed with  more than one ATCA configuration, we
combined   the  data from   different  runs  after  calibration  and continuum
subtraction.    The  final   cubes    were  made with   natural,    uniform or
robust--Briggs' weighting.  Table~3 lists the corresponding rms noise and size
of the restoring beams.

The moment  analysis of the  data cube was  done using GIPSY  (Allen, Ekers \&
Terlouw 1985).  The total intensity  of the \HI\ emission  were derived from a
data cube produced  by smoothing spatially  the original cube to a  resolution
about twice lower than the original.  This smoothed cube  was used as mask for
the original cube:  pixels with signal   below $3\sigma$ in  the smoothed cube
were set to zero in the original cube (van Gorkom \& Ekers 1989).

\HI\ was not detected in three of the target galaxies, and for these  galaxies
the upper limits    on \HI\  emission  were calculated   as  three  times  the
statistical  error on having  no signal over  50 independent velocity elements
(corresponding to 600 km/s) and over four synthesised beams.

The continuum data   were taken from the  emission-free  channels in  the line
data.    They  were also  reduced using   MIRIAD,   and Table~4 summarizes the
results.  The final continuum images were made with uniform weighting.

\section{Results}

The  results  of our \HI\ and  radio  continuum observations are summarized in
Table~5.  In \HI, we detected two of the five  galaxies observed, and three of
the five were detected as  continuum sources. We now give  some details of the
individual galaxies.

\subsection {NGC~1947}

NGC  1947 is  an E1   galaxy with a    system of  at   least three  concentric
minor--axis dust lanes (Bertola et al.\ 1992).  The galaxy  also has a ring of
CO emission (Sage  \& Galletta 1993) and ionized  gas (M\"ollenhoff 1982) with
the same position angle   as the dust lane.    The stars in  NGC\,1947  rotate
around the galaxy's minor axis, perpendicular to the ionized gas rotation axis
(Bertola et al.\ 1992).  The optical spectrum has  central emission lines with
intensity ratios typical   of  LINERs  (Low-Ionization  Nuclear  Emission-line
Regions; Heckman 1980), and an H$\alpha$+[NII] image (Goudfrooij et al.\ 1994)
shows that the ionized  gas has a regular,  disk--like distribution centred on
the galaxy nucleus.

We  detect $3.4 \cdot 10^8$ \msun\  of \HI\ in  NGC 1947,  very similar to the
derived   H$_2$ mass.  The  \HI\ emission  spans a   wide velocity range (from
$\sim$$900$ \kms\  to  $\sim$$1450$  \kms)  with  a width (at   20\%-level) of
$\sim$$550$  \kms.   The  corresponding  value for   $M_{\matHI}/L_B$ is 0.019
$M_\odot/L_{B,\odot}$. The  velocity centroid of  the \HI\  is at $\sim$$1180$
\kms, close to the optically--measured value in Table 1.  The neutral hydrogen
emission is extended and elongated along P.A.$\sim$$127^\circ$, i.e.\ the same
position angle as the dust lane, as can be seen from Fig.~1, where contours in
total  \HI\ are superimposed  on a deep optical   image kindly provided by D.\
Malin.  Although the \HI\ signal is quite weak  (the peak in  the cube is only
$\sim $5 mJy beam$^{-1}$), it is clear that the  distribution of \HI\ is quite
regular.  The \HI\ is not only aligned with  the dust lane  but rotates in the
same sense  and with the same  amplitude as the  CO.  The  total extent of the
\HI\ is $\sim$$2^{\prime}$ ($\sim$13 kpc) in radius, i.e.\ $\sim$$6.2$ times
the optical half-light radius $R_{\rm eff}$.

Position--velocity  plots  along the  major  axis  of   the \HI\  distribution
(Fig.~2a), and along PA = 90$^\circ$ (Fig.\ 2b) confirm that  the \HI\ is in a
warped rotating disk--like  structure  associated  with the  dust   lane.  The
velocity   increases  rapidly  in the inner   part  ($<  30$   arcsec) with  a
`discontinuity' observed in the centre, but  the projected velocity appears to
decrease  in the  outer parts.   The   observed morphology of  the  dust lanes
suggests that the \HI\ might also lie in three or more concentric rings.  This
may   explain   the `discontinuity' seen   near    the systemic  velocity. The
combination of a number   of rings, together with  the  fact that  the spatial
resolution of the \HI\ data is  relatively low, can reproduce the distribution
of the \HI\  in the position-velocity plot.  The  morphology of the dust lanes
suggests that the whole   structure is probably   warped, and the  decrease in
projected velocity at large radii is most likely due to this warp.

The average \HI\ surface density is 0.3 \msun\ pc$^{-2}$. The distribution
peaks in the centre and it is likely that in the centre the \HI\ surface
density is high enough for star formation to occur.  In the radio continuum,
we detect a single point source with a flux density of $\sim$13 mJy (at 1416
MHz) coincident with the optical nucleus.  This appears to arise from a
central region of star formation --- the FIR/radio ratio $q$ (see Helou et
al.\ 1985) is 2.25 for NGC 1947, close to the mean values of 2.14 for spiral
galaxies (Helou et al. 1985) and 2.55 for low--luminosity ellipticals (Sadler
et al.\ 2000).  Thus the \HI\ and star--formation properties of NGC~1947
appear to resemble those of the low--luminosity ellipticals observed by Sadler
et al.\ (2000), consistent with its relatively low optical luminosity ($M_B =
-20.2$\ mag).

We also  detected \HI\ emission from a  companion galaxy, ESO~085--GA088, at a
projected separation  of  16 arcmin (100  kpc)   (from NGC~1947.   This is  an
early--type spiral,  probably of similar    type to the Sombrero Galaxy   (D.\
Malin,  priv.\ comm.),  although  much  less luminous  ($M_B  \simeq  -17.5$).
Unfortunately,   our observations of   this  galaxy are  strongly  affected by
primary-beam attenuation, especially on  the northern side. Nevertheless,  our
data show (Fig.\  3) that this  galaxy is quite \HI-rich ($M_{\matHI}/L_B \geq
0.2$), with   a  very extended  \HI\ distribution   ($R_{\matHI}/R_{B,25} \geq
4$). The projected rotation velocity of about 85 \kms\ appears constant out to
the largest measured point, implying  a mass--to--light ratio of $M/L_B \simeq
20/\sin^2 i$ at a radius of $\sim$13 kpc.

\subsection {NGC~3108}

NGC~3108 is an elliptical galaxy with a minor--axis dust lane (see Fig.~4),
that has been studied in detail by Caldwell (1984).  Fig.~5 shows the light
profile measured from a $V$ band image recently obtained at CTIO, with the
surface brightness plotted against the radius $R$ (a straight line would
indicate an exponential profile), against $\log R$, and against $R^{1/4}$ (a
de Vaucouleurs profile).  A de Vaucouleurs profile with an effective radius of
$R_{\rm eff} = 25.9$ arcsec fits the data well, although there is evidence of
a very faint disk perpendicular to the major axis of the main body.  The
surface brightness profile in Fig.~5 differs slightly from that published by
Caldwell (1984) from photographic data (being systematically fainter by about
0.6 mag), probably because of difficulties in calibrating the older
photographic data.

Caldwell (1984) pointed out a misalignment between the main dust lane and the
minor axis of the light distribution.  Fig.~6 shows optical images (taken with
the CTIO 4-m PFCCD in 1984 and with the CTIO 0.9-m telescope in 1998) in which
a model, based on the light profile, for the light distribution has been
subtracted and a second image where the galaxy has been divided by this model.

Two dust lanes can then be seen along with
a possible inner shell.  Fig.\ 6b shows some excess luminous material
that appears to form a counterpart to the large dust lane (already
mentioned in Caldwell 1984).  The dust lane that appears (in projection)
at larger radii is at position angle $\sim$$109^\circ$ while the dust
lane projected against the nucleus has a position angle of roughly
$130^\circ$, which is closer to the position angle of the minor axis of
the galaxy (the position angle of the optical major axis is 45$^\circ$). 
Caldwell (1984) also detected a disk of ionized gas in the inner 10
arcsec. 

Our \HI\ observations show that  NGC~3108 is a  very interesting galaxy.   Not
only do  we detect a  large amount of \HI\  ($4.6 \times 10^9$ \msun, giving a
value of   $M_{\matHI}/L_B \sim  0.093$) but the   neutral hydrogen  is  quite
extended  and very   regularly  distributed and oriented perpendicular to the
optical major axis.  Fig.~4  shows    the  total \HI\
intensity map superposed on  an optical image.    Although the elongated  beam
shape  of  the    radio  observations  has    `thickened' the  apparent   \HI\
distribution, it is clear that the \HI\  is distributed in a regular, rotating
disk--like structure extended $\sim$$2.5^\prime$ ($\sim 39$ kpc, corresponding
to $\sim$5.8 $R_{\rm eff}$) each side.   Figure~7 gives the position-velocity
map  along the main dust  lane  (P.A.\ 109$^\circ$).   The neutral hydrogen is
distributed over a wide ($\sim$600  \kms\ at 20\%-level) velocity range  (from
$\sim$2300 \kms\ to 2900 \kms) with a  systemic heliocentric velocity of $2610
\pm 5$ \kms.  There is a hint in the total \HI\  image that the \HI\ structure
is warped towards larger position angle in the outer parts.

The  regular  structure  of  the  \HI\   emission is   clearly  shown by   the
position-velocity map  in Fig.~7. The   lack of fast   rotating \HI\ near the
centre suggests that the  \HI\ distribution has  a  central hole of roughly  1
arcminute diameter.  In the outer   parts of the   HI distribution, there  are
hints of a  slight increase (of 15--25  \kms) in the projected velocity  (more
clearly on the E side).

Assuming that the \HI\ distribution has a central hole of 1 arcminute
diameter, the average surface density of the \HI\ is about 0.5-1 \msun\
pc$^{-2}$.  The position--velocity map does not suggest that there are large
gradients in the surface density of the \HI. This means that the \HI\ surface
density over most of the disk is probably too low for star formation, although
some star formation could occur in some isolated regions.  The fact that the
\HI\ is aligned with the faint optical disk suggests that some star formation
may have happened in the past in the \HI\
disk.

We obtained two long--slit spectra of NGC~3108 with the CTIO 4m
telescope in 1984 and SIT-vidicon in order to study the kinematics of
the ionised gas in the centre of NGC 3108 (Fig.\ 8).  One spectrum is
along the position angle of the outer dust lane (PA $\sim 109^\circ$),
and the other perpendicular to the major axis of the light distribution
(PA $\sim 130^\circ$).  The wavelength region covered included the
H$\alpha $ and [NII] emission lines.  The derived rotation curves show
that the small inner disk of ionised gas has very similar kinematics to
the larger \HI\ disk and that both the neutral and the ionised gas have
similar rotation velocities.  It therefore appears that NGC~3108 has a
single gas disk which is mostly ionised in the central regions, and
becomes mainly neutral beyond 30~arcsec radius. 

The gas disk is slightly warped.  The PA of the cental dust lane is
around 130$^\circ$, while the main dust lane and its associated \HI\
disk have a P.A.\ of about 109$^\circ$.  The total \HI\ image suggests
that the P.A.\ increases again in the very outer regions.  The
position--velocity map suggests that the projected rotation velocities
rise slightly in the outer part, probably due to warping of the outer
\HI\ disk to higher inclination.  Assuming that the outermost \HI\ disk
is perfectly edge--on, and that the rotation curve is perfectly flat, an
increase of roughly 20 \kms\ would imply an inclination of about
70$^\circ$ for the inner \HI\ disk.  This is probably just consistent
with the inclination of 75$^\circ$ for the main dust lane as derived by
Caldwell (1984).  It does suggest that the rotation curve of NGC~3108
along its minor axis stays flat out to at least 6 effective radii.

The projected rotation velocity is quite constant at roughly 265 \kms\ from a
radius of 5\,arcsec to the very outer regions of the \HI\ disk. Even without
any detailed modeling it is clear that the rotation curve in this galaxy along
its minor axis is quite different from what expected form the light
distribution (basically a $R^{1/4}$-law), an indication of a dark matter halo
in NGC 3108.  With the approximation of a spherical mass distribution and
circular orbits for the gas, we can make a reasonable estimate for the total
mass of NGC 3108. Taking a de-projected (for an inclination of 70$^\circ$)
rotation velocity of 290 \kms\ at $170^{\prime\prime}$, we derive a value for
the mass-to-light ratio $M/L_B$ of $\sim$18 $M_\odot/L_{B,\odot}$ at $\sim $6
$R_{\rm eff}$.

No radio continuum emission was detected from NGC~3108, with a 3$\sigma$ upper
limit of $\sim $1.4 mJy at 1.4\,GHz. This implies an upper limit to the star
formation rate of 1 $M_\odot$ pc$^-2$.

NGC~3108 is part of a small  group of galaxies,  and we have detected three of
them  in   \HI.   Two,  the  Sbc   galaxy   IC~2539  and an   edge--on  spiral
ESO~435--G028, have already been detected in \HI\ by  Theureau et al.\ (1998).
The third is faint, low surface  brightness an anonymous galaxy (A1000--31) at
$\alpha = 10^{\rm h} 00^{\rm m} 58^{\rm s}$, $\delta = -31^{\circ} 24^{\prime}
12^{\prime\prime}$ (J2000) at a  heliocentric velocity of $\sim$2820 \kms\ for
which we find $M_{\matHI}=6.7\cdot 10^8$ $M_\odot$  and A1000--31 appears to be
very hydrogen rich (see Fig.\ 9).

\subsection{ESO 263$-$G48}

This  is a major--axis  dust--lane elliptical  galaxy for  which the IRAS data
suggests an unusually high dust mass: $M_{\rm  dust} = 9.8 \times 10^6$ \msun. 
Surprisingly, no \HI\ was detected in association with this galaxy in our ATCA
data, and our upper limit is $M_{\matHI} < 0.9 \times 10^8$ \msun).  From this
we derive a tight upper limit to the ratio of the \HI\ mass to blue luminosity
of $M_{\matHI}/L_B < 0.0008$. From the upper  limit on $M_{\matHI}$, we obtain
a value of $M_{\matHI}/M_{\rm dust} < 10$).  Typical values for this ratio are
from a few  hundred to over 1000  (\bf  ref\rm), so the  neutral-to-dust ratio
seems  to be very low in  this galaxy, although  a few more galaxies are known
with similarly  low values for $M_{\matHI}/M_{\rm dust}   $. Inspection of the
IRAS  data shows that  perhaps the FIR fluxes  are  somewhat uncertain because
large complexes of FIR cirrus are found right near ESO 263$-$G48 and confusion
could play some role. Still, a clear point source can be seen  in the FIR data
and  ESO 263$-$G48 most likely   is poor in  atomic hydrogen  compared to what
expected from  the FIR fluxes.

We  do   detect $\sim$$6  \times   10^8$ \msun\  of   \HI\ from  a low--surface
brightness  companion   galaxy (see Fig.\  10),  ESO~263--G47, at a projected
distance of 120 kpc, with a heliocentric velocity of $2396\pm15$ \kms.

In the radio continuum, we detected a single point source with a 1.4\,GHz flux
density  of $\sim$22 mJy coincident with  the optical nucleus of ESO~263--G48.
Radio  continuum  measurements of  this galaxy have   previously  been made at
$\sim$1 arcsec resolution with the VLA by Sadler  et al.\ (1989), who detected
a source 15 arcsec  in size, extended along PA  100 (i.e. nearly perpendicular
to the dust  lane), and with a  flux density of 13.0  mJy at 5 GHz. They  also
measured a core  flux  density of 1.5  mJy  at 5  GHz  within the  central 1.0
arcsec.  The 1.4 GHz continuum flux density is  roughly what would be expected
from the  radio--FIR correlation (e.g.\ Wunderlich  et al.\ 1987) based on the
IRAS flux of ESO 263--G48, but the radio morphology of a linear source centred
on the nucleus and extending perpendicular  to the dust lane  (also seen in an
unpublished 1.4 GHz VLA  image by Laing \&  Kotanyi, private communication) is
much more suggestive of an active nucleus with jets.  The  nature of the radio
continuum emission from ESO 263--G48 therefore remains unclear.

\subsection{NGC~7049}

NGC~7049 is  an S0 galaxy with  a major--axis dust lane,  and is undetected in
\HI\  ($M_{\matHI} <     0.7  \times   10^8$\msun),  from    which we    derive
$M_{\matHI}/L_B <  0.001$.  The galaxy   is detected by  IRAS  and, as for ESO
263$-$G48, we have a  low  upper limit to  the  atomic gas--to--dust ratio  of
$M_{\matHI}/M_{\rm dust} < 29$), which is far lower  than the typical value of
several hundred to a thousand seen in early--type galaxies.

In the radio continuum, we  detect a point  source coincident with the nucleus
of NGC~7049 with flux density  of $\sim 49$ mJy at  1.4\,GHz.  The galaxy  was
also  detected  as a  continuum source  by   Sadler (1984), who measured  flux
densities of 75 and 35 mJy with the Parkes telescope at frequencies of 2.7 and
5.0 GHz respectively.  Slee et al.\ (1994)  detected a compact, flat--spectrum
radio core  with flux density $\sim$15 mJy  at 0.03 arcsec resolution with the
Parkes-Tidbinbilla Interferometer at 2.3  GHz, implying that NGC~7049 hosts an
active nucleus.

We detect \HI\ emission from a companion galaxy in the field of NGC\,7049, the
spiral galaxy ESO~235-G85,  which is about 7\,arcmin  away (90 kpc)  and has a
heliocentric velocity of $2024\pm20$\kms (see Fig.~11).

\subsection{ NGC~7070A}

NGC~7070A  is  an  S0/a  galaxy with  a  skewed   dust  lane  intermediate  in
orientation between the major  and minor axes.  It  was undetected in our ATCA
observations, both in \HI\  and in the radio continuum.   We measure an  upper
limit to the neutral  hydrogen mass of  $M_{\matHI} < 0.5 \times  10^8$ \msun,
giving  $M_{\matHI}/L_B < 0.003$.  As with  the  other two dust--lane galaxies
undetected  in \HI, we  find a low  upper  limit for  the atomic gas--to--dust
ratio: $M_{\matHI}/M_{\rm dust} < 72$.

We  do detect \HI\  emission from the  nearby spiral galaxy NGC~7070.  Fig.~12
shows  the total  \HI\ image  for this galaxy,  and we  measure a heliocentric
velocity of $2450\pm20$ \kms.

\section{Other detections of \HI\ in dust--lane galaxies}

In the literature we have found four more dust--lane ellipticals for which
\HI\ imaging data are available.  Two of them have been observed by us 
(NGC~5266; Morganti   et  al.\ 1997  and IC~5063;  Morganti  et  al.\  1998,
Oosterloo et   al.\ 2000), and  the  other  two are  the  well--known galaxies
NGC~5128 (Centaurus ~A;  van Gorkom et al.\ 1990,  Schiminovich et al.\  1994)
and  NGC~1052 (van Gorkom   et al.\ 1986).   We  summarise the data  for these
galaxies in Table~6, and discuss their characteristics briefly here.

NGC~5266 is   a minor--axis, dust--lane  elliptical galaxy  with  a very large
amount of \HI,  about $2.4 \times  10^{10}$  $M_\odot$, giving $M_{\matHI}/L_B
\sim  0.18$.  The gas extends  to $\sim$$8^\prime$ ($\sim$140  kpc) in radius.
Most of the \HI\ lies almost orthogonal to the optical  dust lane (Morganti et
al.\ 1997) and only a small fraction of the  \HI\ is associated with the dust
lane.  The  inner half of the    \HI\ appears to  be in   a reasonably regular
structure, while the  \HI\ more distant  from the centre  is probably not  yet
settled. A value  of $M/L_B \sim 8$  $M_\odot/L_{B,\odot}$ at $\sim$4  $R_{\rm
eff}$ was found.

IC~5063 (PKS~2048$-$572) is an early--type galaxy hosting a Seyfert 2 nucleus,
and with a dust lane along the major  axis.  It also has  a high \HI\ content:
$8.4 \times 10^{9}$$ M_\odot$ ($M_{\matHI}/L_{B}  = 0.13$).  The \HI\ emission
is elongated ($\sim$38 kpc each side) in the direction of the dust lane and of
the ionized gas disk (Morganti  et al.\ 1998).  The \HI\  is distributed in  a
regularly rotating disk with a slight warp. If we assume circular orbits and a
spherical  mass  distribution,  we  find $M/L_B$   to be $\sim$14   $M_\odot /
L_{B,\odot}$ at about 5.4 $R_{\rm eff}$.

In  NGC~5128, the neutral hydrogen appears  to  be largely associated with the
dust lane  (van Gorkom et  al.\ 1990).  In the inner  part the \HI\ velocities
match the optical emission-line  velocities while in  the outer parts  the gas
disk seems to be not settled yet into  stable orbits.  Recently, \HI\ emission
has also been detected further out in this galaxy,  at distances up to 15\,kpc
from the nucleus (Schiminovich et  al.\ 1994).  This  outer \HI\ appears to be
in a partial ring  inclined  respect to the  dust  lane, rotating in the  same
sense as the main body of the galaxy.

NGC~1052 is classified as an E3 galaxy with a minor--axis dust lane.  \HI\ was
first detected by Knapp et al.\ (1978),  and its distribution has been studied
in detail by van Gorkom et al.\ (1986).  They found  the \HI\ distributed over
about three times the size of the optical galaxy  and roughly perpendicular to
the major axis  (i.e.\ roughly aligned  with the dust lane).  The distribution
of the \HI\ is  not very regular especially in  the outer regions  where tidal
tails  are observed.  The ionized gas  detected in  this galaxy is distributed
roughly in a disk that may not be filled (Davies \& Illingworth 1986).  A more
detailed study (Plana \& Boulesteix 1996)  has shown that two counter-rotating
gaseous components are  actually present  with  an apparent major axis  nearly
perpendicular to  the stellar one, i.e.\ similar  to the position angle of the
neutral hydrogen.  In particular, one of the components has kinematics similar
to the \HI.

NGC 5363 is another early-type galaxy with dust lanes in which HI has been
detected (e.g.\ Thuan \& Wadiak 1982). However, only single-dish data are
available and little is known about the distribution an dkinematics of the
\HI.

\section{Discussion}

We have detected and imaged the neutral hydrogen in two of the five dust--lane
early-type galaxies observed and are able to put sensitive upper limits on the
\HI\ content of the remaining three galaxies.  An interesting result is the
large amount of \HI\ in one of the objects, NGC~3108.  With $4.6 \times 10^9$
\msun\ of neutral hydrogen, NGC~3108 has a value of $M_{\matHI}/L_B = 0.09$
which is at the gas-rich end of the distribution characteristic of ellipticals
(Knapp et al.\ 1985) and is similar to that found in many luminous spiral
galaxies.  The \HI\ in NGC 3108 is distributed in a very regular disk that is
quite large ($\sim$40 kpc radius), and appears to be only slightly warped.
The \HI\ disk appears to have a central hole that is filled by a disk of
diffuse emission from ionized gas.  The average \HI\ surface density (0.5-1
\msun\ pc$^{-2}$) is too low for large-scale star formation.

The other galaxy (NGC~1947) where \HI\ is detected is less gas--rich with
$M_{\matHI}/L_B \sim 0.019$, which is a more typical value for those
elliptical galaxies that are detected in \HI. As in NGC 3108, the \HI\ in NGC
1947 extends to several effective radii and appears to be a disk that, in
contrast to the disk in NGC 3108, is strongly warped.

\subsection{ISM content}

The fact that we do not detect the other three galaxies in \HI\ means that
there is a large range in ISM properties in dust-lane ellipticals, as there is
for early-type galaxies in general. This is the case for the \HI\ content, but
also in the relative content of different indicators of a cold ISM. The upper
limits for $M_{\matHI} / M_{\rm dust}$ for the non-detections are quite
strict, considering that typical values for this ratio (and as observed in NGC
1947 and NGC 3108) are between a few hundred to a few thousand, although a few
more galaxies with very low values exist (Henkel \& Wiklind, 1997). It is
quite unlikely that $M_{\matHI} / M_{\rm dust}$ is so low due to  most of the
\HI\ being ionized,  since such large amounts of ionized gas would be detected
easily. 
A possibility would be
that much of the gas in these galaxies is in molecular (H$_2$) rather than
atomic form. An example of such a galaxy is NGC 759 where $3.5
\times 10^6$ $M_\odot$ of dust is detected (Wiklind et al.\ 1997) while a
5-$\sigma$ upper limit of $10^8$ $M_\odot$ exists for the amount of \HI\ in
this galaxy (Oosterloo, Morganti \& Wiklind, in prep.). This means that for
NGC 759 $M_{\matHI} / M_{\rm dust} < 30$. However, $2.4 \times 10^9$ $M_\odot$
of molecular hydrogen is detected in this galaxy (Wiklind et al.\ 1997).  This
means that $M_{\rm gas} / M_{\rm dust} \sim 700$, not outside the normal range
of values for this ratio (Henkel \& Wiklind, 1997). It also means that all the
cold gas in NGC 759 is in molecular form. The molecular hydrogen is found to
be very centrally concentrated in NGC 759 (Wiklind et al.\ 1997). Perhaps that
in NGC 759 the atomic gas, that is normally more extended than the molecular
gas, has been removed by interactions with neighbouring galaxies, leaving NGC
759 poor in \HI, but not poor in molecular gas. NGC 759 is indeed in a denser
environment that that of galaxies like NGC 3108 or NGC 5266.

Little data on the molecular gas content of these galaxies is available,
although the work of Lees et al.\ (1991) suggests that the molecular gas mass
is typically similar to the \HI\ mass, although also here a large scatter
exist. As noted in Table 1, NGC~1947 has been detected in CO (Sage \& Galletta
1993) with an implied H$_2$ mass of $3.9 \times 10^8$ $M_\odot$, which is
slightly higher than our measured \HI\ mass.  Clearly, we need more sensitive
CO observations of these galaxies to clarify the state of their ISM.  The
158$\mu$m line of [CII] detected by ISO (e.g.\ Malhotra et al.\ 1999) has the
potential to provide an independent and very sensitive probe of the cool ISM
in early--type galaxies, but only a few observations are available so far.

\subsection{Origin  of the \HI\ disks/rings}

It is generally assumed that the neutral hydrogen in ellipticals is related to
a recent merger- or accretion event.  In most numerical simulations of merging
galaxies of similar size, most of the gas quickly falls towards the centre
where it fuels a burst of star formation.  The large, regular \HI\ structures
seen in some of the galaxies discussed here (in particular the very regular
disk in NGC 3108), the fact that in some galaxies these disks exist from the
very centre out to radii of several tens of kpc, and the large amount of \HI\
present in a few galaxies, may seem incompatible with this.  Perhaps if the
\HI\ disks are not formed in a merger of similarly sized galaxies, but instead
are the result of several smaller accretions that, since such accretions are
less ``violent'', can lead to the formation of a large, regular disk in a way
that is to some extent similar to how disks in spiral galaxies form.  The
difference would then be that the surface density in the gas disks in the
early-type galaxies is too low for a significant optical disk to form.  In a
way, they could be ``disk-less disk galaxies'', or perhaps parallels exist
with low surface brightness spiral galaxies.  An alternative is that for a
relatively uncommon set of initial conditions of a major merger, a large
fraction of the gas forms large gas disk.

One of the main arguments found in the literature for a merger related origin
is that the amount of \HI\ is uncorrelated with the optical luminosity of the
galaxy, in contrast to the situation for spiral galaxies (e.g. Knapp et al.\
1985, Roberts \& Haynes 1994).  In the dust--lane ellipticals discussed here,
there is indeed a large scatter in relative \HI\ content, ranging from
\HI--rich objects like NGC 5266 (with $M_{\matHI}/L_B = 0.18$) to very
\HI--poor galaxies like ESO 263--G48 ($M_{\matHI}/L_B < 0.0008$) and NGC 7049
($M_{\rm HI}/L_B < 0.001$).  There also seems to be a large range in the
conditions of the ISM, considering the large range in relative amounts of
different constituents of the cold ISM.

Optical data  suggest that in several of the galaxies discussed here a
significant accretion event must have taken place and that the disk is not a
result of a slower buildup. NGC 5266 has a very low-surface brightness,
extended and irregularly shaped optical counterpart to the large-scale \HI\ 
disk that only shows up at large radius after special data processing
techniques (see Fig.\ 5 in Morganti et al.\ 1997).  In IC 5063, a deep image
(Danziger et al.\ 1981) shows a similar faint, irregular large-scale optical
structure with some indications of optical shells. In NGC 5128 optical shells
are seen at large radius and an association of these shells with the \HI\ at
large radius has been suggested (Schiminovich et al.\ 1994).  The
morphology of these very faint optical counterparts to the outer \HI\ disks
indicates that they are structures that are not settled yet and may have
formed from stellar material from the progenitors that also provided the \HI.

Numerical studies show that a large fraction of the tidal tails formed during
the interaction and merger will remain bound to the remnant (Barnes \&
Hernquist 1996 and references therein).  This gas will slowly `fall back' to
the main body of the galaxy over a long period of time. For the galaxy NGC
7252, Hibbard \& Mihos (1995) predict that about $10^9 M_\odot$ of \HI\ will
return within 15-45 kpc in about 3 Gyr, if the gas structures at large radius
are not disrupted by companions in the mean time. Not all of this gas may not
fall into the centre. Instead, it may form an extended, low surface--density
disk. In a crowded environment, this infall can be disrupted and a disk will
not form.  The regular kinematics of most of the \HI, but the tail-like
features at large radius in NGC 5266 would suggest that this galaxy represents
an intermediately evolved result of such a major merger. IC 5063, with a more
regular \HI\ distribution, could represent an even older merger remnant.  The
very regular appearance of the disk in NGC 3108, combined with the fact that
it goes from centre out to $r=40$ kpc, would mean that, if the disk in NGC
3108 is also due to a single event, it must be an even older merger remnant
than IC 5063. The outer isophotes of NGC 3108 are quite boxy, indicating a
merger of some size has taken place.

Given the rarity of merger remnants with such a large, regular \HI\ disk, the
conditions that gave rise to such a remnants do no occur every
often.Nevertheless, it appears that the formation of large gas disks
is a rare, but natural result of the way early-type galaxies form.

In NGC 1947 no extended low-surface brightness structure is seen in the
optical (D.\ Malin priv.\ comm.).  The amount of gas in NGC 1947 and its lower
$M_{\matHI}/L_B$ may indicate that the \HI\ in this galaxy could result from a
single `small' accretion, i.e.\ an accretion with a mass ratio of more than a
few.

\subsection{Evolution of the HI disks}

The surface densities of the \HI\ in most of the galaxies are quite low. For
NGC 3108 we find that it must be in the range 0.5-1 \msun\ pc$^{-2}$ over most
of the disk. For NGC 1947 we find that the average surface density is 0.3
\msun\ pc$^{-2}$ but the value for the central region is probably higher.
These surface densities are quite low and, except for the central regions, no
large-scale star formation will occur in these galaxies. Not much evolution
will happen in these disks at large radius. Even though the galaxies discussed
here are early-type galaxies, some of these galaxies will remain gas rich for
a very long period of time and the gas should be considered a fundamental
constituent of the galaxy.

It is worth remarking on the difference  between the morphology of the neutral
hydrogen in {\sl luminous} dust--lane ellipticals, like NGC 3108 and NGC 5266,
and in    low--luminosity  E/S0 galaxies.  In   the  low--luminosity galaxies,
regular disks of \HI, extending out to a few  effective radii, are very common
(Lake et al.\ 1987;  Sadler et al.\ 2000).  However,  the \HI\ distribution is
strongly peaked towards the  centre, in contrast to  what  we see in the  more
luminous dust-lane ellipticals, and   the  central \HI\ surface   densities in
these  low-luminosity galaxies are high enough  for significant star formation
to  occur in their centres.    Several of  the luminous dust-lane  ellipticals
considered here have ionized gas in the central region in the form of disks or
rings that, in  some cases like NGC 3108  are inner extensions of  the neutral
gas disks.  The optical spectra of these  central ionized gas disks  (with the
exception of IC 5063,  which is a  Seyfert 2 galaxy) are  typically LINER-like
and not typical of star  forming regions.  The ionization  mechanism of gas in
LINERs is still  unclear (Colina \& Koratkar  1997), but is usually attributed
either to shocks (Heckman 1980, Dopita \& Sutherland 1995), old UV-bright post
AGB stars (Binette et al.  1994) hot high-metallicity O-type stars (Filippenko
\&  Terlevich  1992)  or   to thermal  interaction   with a   hot  ISM  (e.g.\ 
Goudfrooij 1998). Whatever  the mechanism,    it  appears that the conditions    in
luminous  early-type  galaxies,  in   contrast  to  less   luminous early-type
galaxies, are such that the hydrogen near the centre cannot remain neutral but
becomes ionized.

\subsection {Dark matter}

Some of the  regular \HI\ structures can be  used for studying the dark matter
properties at  large radii.  In general, a   complication is that it  is often
difficult to determine the inclination of the \HI\ disk in early-type galaxies
(e.g.\ Sadler et  al.\ 2000). For dust--lane  galaxies, however, the situation
is more promising since the inclination of the \HI\  structures can be assumed
to be close to  edge--on and small uncertainties  in the inclination angle  do
not have a large effect on the  derived total mass.  In  NGC 3108 the rotation
curve is basically flat,  as it is in luminous  spirals. If we assume circular
orbits and a spherical mass distribution, we derive a value of $M/L_B \sim 18$
$M_\odot/L_{B,\odot}$ at 6 $R_{\rm eff}$ for NGC~3108. This compares well with
values of 15--20 $M_\odot/L_{B,\odot}$  that are typically found in  elliptical
galaxies at radii of 6-7 $R_{\rm eff}$ using \HI\ data (e.g.\ Morganti et al.\
1997; Morganti et al.\ 1998). These numbers imply that significant amounts or
dark matter are present at large radii in these galaxies.

It is  interesting to compare  the mass--to--light  ratio  values derived from
neutral hydrogen data with those obtained from X--ray observations, since very
different tracers of   the potential are  used.   From X--ray  data, values of
$M/L_V \sim 25 h_{80} M_\odot/L_{V,\odot}$ within 6  $R_{\rm eff}$ are usually
found   (e.g.\    Loewenstein 1998)  which     converts  to $M/L_B    \sim  16
M_\odot/L_{B,\odot}$ for the value of the  Hubble constant used  by us in this
paper.  Although both methods have  their uncertainties,  there appears to  be
good agreement  between  the \HI\  and   X--ray estimates  of  the  masses  of
early--type galaxies.

\section {Conclusions }

We have presented new \HI\ observation of five  dust--lane ellipticals. In two
galaxies we detect \HI\ emission.  In both galaxies the \HI\  is in a  warped,
disk-like structure, with (in particular in NGC 3108) regular kinematics.  For
the remaining tree galaxies we derive sensitive upper limits that imply a very
low \HI\ content compared to the dust mass. Differences in merging history may
perhaps explain the large range  in the relative  content of different tracers
of the cold ISM.

By adding data from the literature, we find several dust-lane elliptical
galaxies where the \HI\ is regularly distributed in disks/rings of several
tens of kpc in size, and at least three dust-lane galaxies with more than
$10^9 M_\odot$ of \HI. The data collected suggests that most dust-lane
galaxies probably represent the results of major mergers as seen at a late
stage in their evolution.  In these systems, the late infall of gas has
produced regular, settled \HI\ disks. In some cases, the disk could also be
the result of accretions of several small galaxies, spread out in time.  The
\HI\ surface density is quite low over most of these disks and little or no
recent star formation has occurred (except possibly in the very centre).
These gas disks will evolve very slowly and the galaxies will remain gas rich
for a very long period of time.
The presence of large, regular \HI\ structures suggests that a
significant fraction of elliptical galaxies may have a long--lived cool
interstellar medium.

The regular distribution and kinematics  of the \HI\ in  NGC 3108 allow us  to
derive  the mass--to--light ratio  in this galaxy. We  find a  value of $M/L_B
\sim 18$  $M_\odot / L_{B,\odot}$   at 6 $R_{\rm   eff}$.  This value is  very
similar  to   what is found   in   other elliptical  galaxies  based   on \HI\
data. Estimates of $M/L$ using \HI\  are very similar to  those based on X-ray
data.

\acknowledgements

We would like to thank M.M.  Phillips and R.  Covarrubias for taking the
0.9m image of NGC 3108 and David Malin for kindly providing the deep
optical image of NGC~1947.

\newpage

\begin{center}
{\small
{\bf Table 1.} {Properties of the observed  galaxies}
\\
\smallskip
\begin{tabular}{lcccccc} \hline \hline
&{\bf  NGC\,1947} & {\bf NGC\,3108}& {\bf ESO 263--G48}& {\bf NGC\,7049} &
{\bf NGC\,7070A} & {\bf Refs.}\\
\hline
RA (J2000.0)   &  05 26 47.5 & 10 02 29.4  & 10 31 11.2  & 21 19 00  & 21 31 47.1 
\\
Dec (J2000.0)  & -63 45 40   & -31 40 37   &-46 15 02    & -48 33 43 & -42 50 45  
\\
Type of dust lane &  minor      & minor       &  major     &   major       &  skewed   \\
B$_{\rm T}^{\ 0}$&   11.50   & 12.31       &  11.69     &  11.57     & 
13.32$^b$ & a \\
r$_{\rm eff}$ (arcsec) &     19.0    & 25.9        &   ...       &   ...       &   ...     \\
v$_h\odot$ (km/s)   &  1100       &  2673       &  2889      &  2231      &  2391   \\
Distance (Mpc) &   22.0      & 53.5        &   57.8     &  44.6      &  47.8   \\
Ang. scale (kpc/arcsec) & 0.107 & 0.259 &  0.280 & 0.216 & 0.232  \\ 
$M_{B}$ (mag)  &   --20.21   & --21.33     &  --22.14   &  --21.68   & --20.08  \\
$L_{B}$ ($10^{10}$\lsun) & 1.8 & 5.0     &  10.5     &  6.8      &  1.6   \\
F(60$\mu m$) (Jy) & 1.05      & $<$0.12        &  1.2       &  0.53      &
0.23 & c     \\
F(100$\mu m$)(Jy) & 4.14      & 0.73        &  4.4       &  1.93      & 0.66
& c   \\
T$_{60/100\mu m}$   &  28.3       & $<$23.8        &  29.1      &  29.3      & 31.8    \\
M$_{\rm dust}$ ($10^{5}$\msun) & 15.3 & 42.3   &  97.8    &  25.1      &  6.6    \\
$L_{\rm FIR}$ ($10^{9}$\lsun) & 1.21 & 1.08 &  9.13     &  2.40      &  1.04
&  \\
M$_{\rm H2}$ ($10^{8}$\msun) & 3.92 &  ...     &    ...    &  ...    & 
... & d \\
    
\hline
\hline
\end{tabular}
$^a$ Revised RC3 (de Vaucouleurs et al. 1991)\\
$^b$ for NGC~7070A  we give B$_{\rm T}$\\
$^c$ Knapp et al. (1989)\\
$^d$ Sage \& Galletta (1993)
}

\end{center}

\newpage
\begin{center}
{\bf Table 2.} {Log of ATCA Observations}

\smallskip  

\begin{tabular}{cccclc} \hline \hline
{Galaxy} &{Date} &   {Configuration} & {Baseline} &  {Frequency} & {Time} \\
         &       &                   & {range} (m)&   (MHz)      &     \\
 \hline
 NGC\,1947 & Sep96   & 375   &  31--459      & 1416        & 12h\\
           & Sep95   & 750D  &  31--719      & 1416        & 12h\\
           & May95   & 1.5D  &  107--1439    & 1416        & 12h\\
 NGC\,3108 & Apr96   & 375   &  31--459      & 1407        & 12h \\
           & Nov96   & 750A  &  77--735      & 1407        & 12h \\
 ESO~263--G48 & Apr96  & 375 &  31--459      & 1407        & 12h \\
 NGC~7049  & Sep97     & 375 &  31--459      & 1409        & 12h \\
 NGC~7070A & Sep96     & 375 &  31--459      & 1409        & 6h \\
           & Sep97     & 375 &  31--459      & 1409        & 6h \\
\hline
\hline
\end{tabular}
\end{center}

\bigskip

{\small
\begin{center}
{\bf Table 3.} {Instrumental Parameters: \HI\ observations}

\smallskip

\begin{tabular}{lccccc} \hline \hline
 & {\bf NGC\,1947} & {\bf NGC~3108} &{\bf ESO~263--G48} &{\bf NGC~7049} & {\bf
NGC~7070A} \\
\hline
Field Center (J2000) &         &           &       &      &   \\
RA                   & 05 26 47 & 10 02 29  & 10 31 11 &  21 19 00 & 21 31 47 \\
Dec                  & --63 45 40& -31 41 01 & --46 15 16  & --48 34 19  & --42 51 47 \\
Synthesized beam &
$55^{\prime\prime} \times 52^{\prime\prime}$ &
$135^{\prime\prime} \times 57^{\prime\prime}$ &
$160^{\prime\prime} \times 98^{\prime\prime}$ &
$148^{\prime\prime} \times 98^{\prime\prime}$ &
$170^{\prime\prime} \times 92^{\prime\prime}$ \\
 & & & & &\\
Beam p.a.   & $5^\circ$ & $7^\circ$ & $1^\circ$ & $-1^\circ$ & $0.9^\circ$  \\
rms noise in channel maps    & 0.9   &  0.9   & 1.15 & 1.48 & 0.83  \\
(\mJybeam)   &       &        &     &     \\
\hline
\hline
\end{tabular}

\newpage

{\bf Table 4.} {Instrumental parameters: Radio continuum observations}
\\
\smallskip
\begin{tabular}{lcccc} \hline \hline

Galaxy & Array  & Beam, PA &  rms  \\
       &        &  (arcsec) & (\mJybeam)    \\
\hline
NGC\,1947   & 375  &    $80\times 71$(8$^\circ$)    &  0.40    \\
NGC\,3108   & 750  &    $90\times 37$(13$^\circ$)  &  0.45   \\
ESO~263--G48& 375  &    $93\times 68$(0$^\circ$)    & 0.35     \\
NGC\,7049   & 375  &    $93\times 68$(-4$^\circ$)   & 0.15     \\
NGC\,7070A  & 375  &    $100\times 92$(4$^\circ$)   & 0.43   \\
\hline
\hline
\end{tabular}
\end{center}

}

\begin{center}
{\small

{\bf Table 5.} {Properties of the observed  galaxies}
\\
\smallskip
\begin{tabular}{lccccc} \hline \hline
&{\bf  NGC\,1947} & {\bf NGC\,3108}& {\bf ESO 263--G48}& {\bf NGC\,7049} &
{\bf NGC\,7070A} \\
\hline
M$_{\rm HI}$ ($10^{9}$\msun) & 0.34  &  4.59    &  $<0.09$   &  $<0.07$      & $<0.05$  \\
M$_{\rm HI}$/L$_B$           & 0.019 &  0.092   &  $<0.0008$ &  $<0.001$     & $<0.003$  \\
M$_{\rm HI}$/M$_{\rm dust}$  & 222.9 &  1084.5  &  $<9.6$    &  $<29.1$      & $<71.6$        \\
S$_{\rm 21\,cm}$ (Jy)        & 0.013 & $<0.0014$ &  0.022     & 0.049         & $<$0.0013 \\
log P$_{\rm 21\,cm}$ W/Hz    & 20.95 & $<$20.67 & 21.96      & 22.08         & $<$20.56  \\
\hline
\hline
\end{tabular}

{\bf Table 6.} {Dust-lane galaxies from the literature}
\\
\smallskip
\begin{tabular}{lcccccccccc} \hline \hline

Galaxy & D & $L_B$ & $M_{\rm HI}$ & $\log P_{\rm 21\, cm}$ & $T_{\rm dust}$ &
$M_{\rm dust}$ & $L_{\rm FIR}$ & $M_{\rm H2}$ & $M_{\rm HI}/L_B$
& $M_{\rm HI}/M_{\rm dust}$ \\
 &  Mpc & $10^{10}$ \lsun & $10^9$\msun & W/Hz & K & 10$^5$\msun & 10$^9$\lsun & 10$^8$\msun & & \\
\hline
IC 5063   & 68.0 & 6.3  &  8.4 &  23.87 &  54  &  12.3  &  30.3 &   6.5 & 0.13  & 6798 \\
NGC 5266  & 61.5 & 13.9 & 25.2 &  21.74 &  31  &  61.3  &   9.4 &  22.2 & 0.18  &  380 \\
NGC 5128  &  3.7 & 2.4  &  0.4 &  23.52 &  36  &   7.6  &    10 &     2 & 0.017 &  285 \\
NGC 1052  & 29.4 & 3.6  &  0.46 &  22.95 & 41  &   1.9  &   1.2 &  0.33 & 0.013 & 2455 \\
\hline
\hline
\end{tabular}

}
\end{center}

\newpage

\begin{figure}
%\centerline{\psfig{figure=oosterloo.fig1.ps,width=14cm}}
\caption{Map of total  \HI\ distribution in NGC~1947 superimposed to the
optical image from the DSS. The lowest contour is at  $2.4 \times 10^{19}$
\atms\ while the increment is  $4.7 \times 10^{20}$ \atms.}
\end{figure}

\begin{figure}
%\centerline{\hss\psfig{figure=oosterloo.fig2a.ps,width=7cm,angle=0}
%\psfig{figure=oosterloo.fig2b.ps,width=7cm,angle=0}
%\hss}
\caption{
{\sl left}) Position-velocity maps obtained from a slice along P.A.\ =
127$^\circ$. The contour levels are --1.2, 1.2, 2.4 and 3.6 mJy beam$^{-1}$.
{\sl right}) Position-velocity maps obtained from a slice along P.A.\ =
90$^\circ$ showing that also on the Eastern side the projected velocities go
down. The contour levels are --1.2, 1.2, 2.4 and 3.6 mJy beam$^{-1}$.}
\end{figure}

\begin{figure}
%\centerline{\psfig{figure=oosterloo.fig3.ps,width=14cm,angle=90}}
\caption{Total intensity map of the companion galaxy of NGC~1947, 
ESO~085--GA088}
\end{figure}

\begin{figure}
%\centerline{\psfig{figure=oosterloo.fig4.ps,width=14cm}}
\caption{Map of total  \HI\ distribution in NGC~3108 superimposed to the
optical image from the DSS. The contour levels start at  $2.9 \times 10^{19}$
\atms\ in steps of  $2.9 \times 10^{20}$ \atms.}
\end{figure}

\begin{figure}
%\centerline{\psfig{figure=oosterloo.fig5.ps,width=14cm}}
\caption{The light profile of NGC~3108 based on a V band
image (from CTIO). The surface
brightnesses is plotted against the radius $R$, 
against $\log R$, and against $R^{1/4}$.} 
\end{figure}

\begin{figure}
%\centerline{
%\psfig{figure=oosterloo.fig6a.ps,width=7cm,angle=0}
%\hss
%\psfig{figure=oosterloo.fig6b.ps,width=7cm,angle=0}
%}
\caption{$V$ image of NGC 3108 with a model based on the light profile
subtracted, showing the two dustlanes and a possible shell (left), the same $V$
image divided by the model light distribution, showing the faint luminous ring 
(right)}
\end{figure}

\begin{figure}
%\centerline{\psfig{figure=oosterloo.fig7.ps,width=14cm,angle=0}}

\caption{Position-velocity map of NGC 3108 obtained from  a slice along
P.A. = 109$^\circ$. The contour levels are from --2.7 to 13.5 mJy beam$^{-1}$
(excluding the zero contour) in steps of 1.35 mJy beam$^{-1}$.}
\end{figure}

\begin{figure}
%\centerline{\psfig{figure=oosterloo.fig8.ps,width=14cm,angle=0}}

\caption{Radial velocities with respect to the systemic velocities of the
central ionized gas disk, as derived
from the optical spectra. The arrows indicate the projected velocities of the
\HI\ at larger radii.}
\end{figure}

\begin{figure}
%\centerline{\psfig{figure=oosterloo.fig9.ps,width=14cm}}
\caption{Total intensity map of the faint companion  of NGC~3108.}
\end{figure}%

\begin{figure}
%\centerline{\psfig{figure=oosterloo.fig10.ps,width=14cm,angle=-90}}
\caption{Total intensity map of the companion galaxy of ESO~263--G48.
In the inset is shown a zoom-in optical image of ESO~263--G48 taken from
DSS2. The major axis dust lane is clearly visible.
} 
\end{figure}

\begin{figure}
%\centerline{\psfig{figure=oosterloo.fig11.ps,width=14cm,angle=-90}}
\caption{Total intensity map of the companion galaxy of NGC~7049} 
\end{figure}
\begin{figure}
%\centerline{\psfig{figure=oosterloo.fig12.ps,width=14cm,angle=-90}}
\caption{Total intensity map of the companion galaxy of NGC7070A} 
\end{figure}

\end{document}